\documentclass[aps,amssymb,superscriptaddress,showpacs,prb]{revtex4}
\usepackage{epsfig}
\begin{document}

\title{Non-analytic curvature contributions to solvation free energies:
influence of drying}

\author{R. Evans}
\affiliation{H. H. Wills Physics Laboratory, University of Bristol, 
Bristol BS8 1TL, UK}
\affiliation{Max-Planck-Institut f\"{u}r Metallforschung - Heisenbergstr. 3 
70569 Stuttgart, Germany}
\author{J.R. Henderson}
\affiliation{School of Physics and Astronomy,
University of Leeds, Leeds LS2 9JT, UK}
\author{R. Roth}
\affiliation{Max-Planck-Institut f\"{u}r Metallforschung - Heisenbergstr. 3 
70569 Stuttgart, Germany}
\affiliation{ITAP, Universit{\"a}t Stuttgart - Pfaffenwaldring 57 
70569 Stuttgart, Germany}
\date{\today}

\begin{abstract}
We investigate the solvation of a hard spherical cavity, of radius $R$,
immersed in a fluid for which the interparticle forces are short ranged. For
thermodynamic states lying close to the liquid binodal, where the chemical
potential deviation $\delta \mu\equiv \mu - \mu_{co}(T)$ is very small and
positive, complete wetting by gas (drying) occurs and two regimes of
interfacial  behavior can be identified. These are characterized by the
length scale $R_c=2 \gamma_{gl}^\infty/(\Delta \rho \delta \mu)$, where
$\gamma_{gl}^\infty$ is the planar gas-liquid surface tension and $\Delta
\rho$ is the difference in coexisting densities at temperature $T$. For
$R>R_c$, the interfacial free energy and the density profile of the fluid near
the hard wall can be expanded in powers of the curvature $R^{-1}$, in keeping
with the analysis of Stillinger and Cotter, J. Chem. Phys. {\bf 55}, 3449
(1971). In the other regime, $R<R_c$, the interfacial free energy and its
derivatives acquire terms depending on $\ln R$. Since $R_c^{-1}$ can be made
arbitrarily small this implies non-analytic behavior, as $R^{-1}\to 0$, of the
work of formation of a hard spherical cavity and of the Gibbs adsorption and
the fluid density at contact with the wall. Our analysis, which is based on an
effective interfacial Hamiltonian combined with exact statistical mechanical
sum rules, is confirmed fully by the results of microscopic density functional
calculations for a square-well fluid. We discuss the repercussions of our
results for solvation phenomena, emphasizing that non-analytic behavior
equivalent to that we find for complete drying in solvophobic systems will
also arise in the case of complete wetting, i.e. when liquid films are
adsorbed on the surface of large (colloidal) particles or at curved
substrates. We re-assess various results in the important but neglected
Stillinger-Cotter paper, where drying was not considered explicitly, in the
light of our present analysis.
\end{abstract}
\pacs{68.08.Bc,05.70.Np} 

\maketitle

\section{INTRODUCTION}
If a solute is to dissolve in a solvent the latter must pay the price of 
forming cavities capable of accommodating the solute molecules. This process 
can usefully be imagined as being carried out in two steps, with the first
step involving the insertion of a hard cavity in which all attractive
solute-solvent interactions are turned off. Treating the basic model of hard
cavity solvation is the aim of scaled particle theory (SPT).\cite{spt}
One of the key issues in the chemistry and physics of solvation is
understanding the significance of solute geometry and, in particular, the
size-dependence of the free-energy of cavity insertion. If one naively equates
cavity formation with bubble formation, then this issue reduces to
understanding the curvature dependence of the surface tension of a
bubble. Gibbs\cite{gibbs} and later Tolman\cite{tolman} proposed analytic
expansions of the interfacial free-energy in powers of the interfacial
curvatures. This approach was adopted by the founders of SPT and remains today
a commonly-used ansatz for treating inhomogeneous fluid phenomena.\\ 

There is, however, one aspect of hard cavity-solvation that SPT is unable to
describe even qualitatively. Namely, when the radius of curvature is large and
the liquid solvent approaches liquid-gas coexistence, i.e. the chemical
potential $\mu \rightarrow \mu_{co}^+(T)$. In this special case, it is known
from the theory of wetting/drying transitions that a thick film of gas must
appear at the surface of the hard cavity.  More specifically, the planar hard
wall-liquid interface is dry (completely wet by gas) in the limit $\mu
\rightarrow \mu_{co}^+(T)$ for all temperatures $T$ where liquid-gas
coexistence occurs.\cite{drying} It is straightforward to show that standard
SPT is unable to accommodate the drying of a hard cavity in the large radius
limit.\cite{jrhSPT} Recently, Chandler and co-workers have proposed that a
full understanding of the hydrophobic effect (of much importance in biology as
well as chemistry) requires the incorporation of drying phenomena as the size
of hydrophobic solutes increase.\cite{hydro} One of the key aspects of their
interpretation is the realization that at standard temperature and pressure
water is very close to saturation (bubble formation), i.e. the relevant
thermodynamic field (chemical potential or bulk pressure) is very close to its
value at bulk coexistence. Because of the ubiquitous presence of attractive
power-law (dispersion) interactions between solute and solvent true drying
films are not easy to create in the physical world. Even weakly attractive
solute-solvent interactions will lead to partial rather than complete drying
so that the thickness of the region of depleted density is greatly
reduced. Nevertheless, it is argued that the incipient presence of drying
should play an important role in hydrophobic phenomena at large length
scales.\cite{chandler2}\\  

The breakdown of SPT close to saturation implies more than a simple failure to
predict the growth of drying films. Drying at a planar substrate is an example
of interfacial critical phenomena and as such necessarily requires the
existence of non-analytic contributions in, say, $\mu - \mu_{co}(T)$ to the
interfacial free energy.  For a curved substrate one might expect that this in
turn implies non-analytic terms in the curvature expansions of surface tension
and of surface order parameters (derivatives of the interfacial free energy
with respect to thermodynamic fields). Such a scenario calls into question the
whole basis of the Tolman and SPT approach. In fact, the formal status of the
analytic curvature expansion has been controversial in the past, see
e.g. Refs.~\onlinecite{jrhrevtolman,jrhjsr}. Unbeknown to those authors and
apparently to most of the inhomogeneous fluids community, Stillinger and
Cotter\cite{stillcott} had, in 1971, provided what, at first sight, appears to
be a proof that for a general solvent the curvature expansion of the
interfacial free energy is analytic in $R^{-1}$ for spherical hard-body
cavities of radius $R$. Notwithstanding the existence of this proof, the
incorporation of wetting/drying phenomena at curved substrates has led authors
to identify non-analytic $\ln R$ factors in the curvature expansion of the
interfacial free energy and order parameters, such as the adsorption and
contact density, for solvents where the interparticle potential is
finite-ranged.\cite{holyst,evans1,nap} The main purpose of our present paper
is to elucidate the non-analytic behavior identified explicitly in
Ref.~\onlinecite{evans1} and place it in the context of the proof of
Stillinger and Cotter. Although the mathematical details and hence our
discussion pertain directly to complete drying at hard spherical cavities, we
emphasize that the non-analyticities and the cross-over behavior which we
describe below are equally relevant to the physically important case of
wetting films of liquid adsorbed on the surfaces of curved substrates and
colloidal particles.\cite{dietrich1}\\  

We focus, in the main, on model solvents where the interparticle potential is
finite-ranged, say a square-well or a truncated Lennard-Jones fluid.  The
presence of dispersion interactions leads to different non-analyticities which
will be discussed elsewhere.\cite{evans2} Our starting point will be the
phenomenological interface Hamiltonian appropriate to the wetting/drying of
curved substrates, assuming finite-range interactions,\cite{intham} in which
the binding potential is expressed as the excess Grand potential of fluid
surrounding a hard spherical cavity of radius $R$. We consider a reservoir of
liquid at chemical potential $\mu>\mu_{co}(T)$ whose pressure is denoted by
$p$. If $V_{tot}$ is the total system volume including the cavity then the
volume accessible to the fluid is $V_{acc}=V_{tot}-4 \pi R^3/3$. The
wall-fluid boundary of infinite repulsion is located at $r=R$ and this induces
a film of gas to grow out to a radius $R+\ell$ where a fluctuating gas-liquid
interface forms with surface tension $\gamma_{gl}$. The gas film is considered
to have a pressure $p_g$, the pressure of the metastable bulk gas at the same
chemical potential $\mu$, so that its grand potential is $-p_g V_g$, where
$V_g=4 \pi[(R+\ell)^3-R^3]/3$ is the volume of the film. It follows that the
excess grand potential can be expressed as
\begin{eqnarray}
\Omega^{ex}(\ell;T,\mu,R) &\equiv& \Omega + p V_{acc} \nonumber \\
&=& (p-p_g) V_g + 4\pi\left[R^2\gamma_{wg} 
+ (R+\ell)^2\gamma_{gl} + R^2 a(T) \xi \exp\{-\ell /\xi\}\right]\,\,,
\label{eq1}
\end{eqnarray}
where $\gamma_{wg}$ is the surface tension of the wall-gas interface. The
final term in Eq.~(\ref{eq1}), involving the gas-phase bulk correlation length
$\xi$ (the decay length of the gas tail of the planar gas-liquid interface),
is the leading-order mean-field interaction between the gas-liquid interface
and the wall-gas interface. For a hard wall $a(T)>0$ for all temperatures $T$
at which gas-liquid coexistence occurs. From the Gibbs-Duhem equation at fixed
temperature $T$ we can write (defining $\Delta\rho \equiv \rho_l-\rho_g$, the
difference of the coexisting densities)
\begin{equation}
p-p_g = \Delta\rho\delta\mu\,\,,
\label{eq2}
\end{equation}
where $\delta\mu \equiv \mu - \mu_{co}(T)$ denotes the difference between the
chemical potential and its value at bulk saturation. The equilibrium thickness
of the gas film $\ell_{eq}(T,\mu,R)$ follows by minimizing the Grand potential
as a function of $\ell$ at fixed $T,\mu,R$ (and hence at fixed pressures and
surface tensions).  To leading-order in $\ell_{eq}/R$ this yields 
\begin{equation}
\frac{\ell_{eq}}{\xi} = \ln\left\{\frac{aR}{\Delta\rho\delta\mu R + 2\gamma_{gl}
^{\infty}}\right\}\,\,,
\label{eq3}
\end{equation} 
where superscript $\infty$ denotes the planar limit at saturation. Noting that
at worst $\ell_{eq}$ is no bigger than order $\ln R$, we can substitute
(\ref{eq3}) back into Eq.~(\ref{eq1}) to identify the first two terms in a
curvature expansion of the equilibrium interfacial free-energy:
\begin{equation}
\Omega^{ex}_{eq}(R,\mu) = 4\pi R^2\left\{\gamma_{wg}(R,\mu)+\gamma_{gl}(R)+(\xi+\ell_{eq}(R,\mu))
\left[\Delta\rho\delta\mu +\frac{2\gamma_{gl}^{\infty}}{R}\right]\right\}\,\,.
\label{eq4}
\end{equation}
Hereinafter we suppress $T$, since all of our analysis will be isothermal, and
in addition for any specific quantity we list only those variables that
contribute at the order of the expansions discussed below. We also refrain
from constantly reminding the reader of the presence of higher-order terms. To
go beyond our analysis would first require one to reassess the possible need
to include additional terms and/or thermodynamic field variations of the
quantities appearing in the interface binding potential Eq.~(\ref{eq1}). For
example, for our purposes we do not need to distinguish between
$\gamma_{gl}(R+\ell_{eq},\mu)$ and $\gamma_{gl}(R)$.\\ 

The key conclusion of our work is already apparent from the introductory
analysis above. Namely, the presence or absence of non-analytic $\ln R$
contributions is controlled by a length scale 
\begin{equation}
R_c \equiv \frac{2\gamma_{gl}^{\infty}}{\Delta\rho\delta\mu}\,\,, \label{eq5}
\end{equation}
through the asymptotic approach to the planar limit at gas-liquid coexistence
(where both $R$ and $R_c$ are infinite), 
\begin{equation}
\frac{\ell_{eq}}{\xi} \rightarrow 
\left\{
\begin{array}{c}
\ln\left\{\frac{aR}{2\gamma_{gl}^{\infty}}\right\} \,\,,\, R \ll R_c\\
\ln\left\{\frac{a}{\Delta\rho\delta\mu}\right\} \,\,,\, R \gg R_c
\end{array}\,\,.
\right.
\label{eq6}
\end{equation}
The length scale (\ref{eq5}) is identical to the length that controls
capillary evaporation between planar hard walls, or, indeed, capillary
condensation between planar walls that are completely wet.\cite{evanscc}
Chandler and co-workers\cite{hydro} have linked this length scale to the
hydrophobic attraction between large solutes in water; for water at room
temperature and pressure $R_c \simeq 1.4\mu$m, a surprisingly large length
scale.\\  

In the following section we re-analyze the arguments of Stillinger and Cotter,
that might otherwise lead the reader into rejecting outright the presence of
non-analyticity (terms involving $\ln R$) in the curvature expansion of
interfacial free energies and order parameters.\cite{stillcott,reiss} In
Secs.~III-V we detail the cross-over between a non-analytic and an analytic
approach to the planar limit, both for the interfacial free energy and for
interfacial order parameters. Sec. VI presents Density Functional Theory (DFT)
results for a square-well fluid, that illustrate the existence of this
cross-over. We conclude in Sec.~VII with a discussion of the physical
significance of our analysis and comments on some of the additional
interesting material to be found in the paper by Stillinger and Cotter.

\section{STILLINGER AND COTTER SUM RULE ANALYSIS}
There are three exact results that can be applied to a spherical hard cavity
immersed in an arbitrary solvent: 
\begin{eqnarray}
\frac{\partial \Omega^{ex}_{eq}(R,\mu)}{\partial R} &=& 
4\pi R^2 k_BT\left[\rho_w(R,\mu)-\rho_w(\infty,\mu)\right]\label{eq7}\,\,,\\
-\frac{\partial \Omega^{ex}_{eq}(R,\mu)}{\partial \mu} &=& \Gamma(R,\mu) \equiv 
4\pi \int_{R}^{\infty}drr^2\left[\rho(r)-\rho_b(\mu)\right]\label{eq8}\,\,,\\
\frac{\partial \Gamma(R,\mu)}{\partial R} &=& -4\pi R^2 k_BT\frac{\partial}{\partial \mu}
\left[\rho_w(R,\mu)-\rho_w(\infty,\mu)\right]\,\,.
\label{eq9}
\end{eqnarray}
The first two of these results are virial route and compressibility route sum
rules, respectively. A brief review of their derivation and physical content
is given in the appendices to Ref.~\onlinecite{jrhSPT}. The quantity
$\rho_w(R,\mu)$ denotes the limiting value of the density profile $\rho(r)$ at
the surface of the cavity ($r=R$) and $\rho_w(\infty,\mu) = p/k_BT$, where $p$
is the pressure of the reservoir, is its zero curvature (planar)
limit. Eq.~(\ref{eq8}) is the Gibbs adsorption equation for this problem;
$\rho_b(\mu)$ is the reservoir density. The third equation (\ref{eq9}) is the
Maxwell relation that follows directly from the previous two sum rules. It is
equivalent to the derivative w.r.t. $R$ of Eq.~(4.13) of Stillinger and
Cotter,\cite{stillcott} although the latter was derived by a much less direct
route. Note that Eq.~({\ref{eq9}) is interesting in that it relates the
derivative w.r.t. the field $\mu$ of a {\em local} quantity $\rho_w$ to the
derivative w.r.t. $R$ of an {\em integrated} quantity, the Gibbs adsorption
$\Gamma$. Stillinger and Cotter consider the consequences of the assumption of
an analytic curvature expansion of the adsorption $\Gamma$ and the contact
density $\rho_w$, in the context of the Maxwell relation (\ref{eq9}). We
re-visit their argument. Let us introduce the variable $z\equiv r-R$ and
assume following Ref.~\onlinecite{stillcott} an analytic expansion of the
density profile, thus
\begin{eqnarray}
\rho(z;R) = \rho(z;\infty) + \frac{\rho'(z)}{R} + \frac{1}{2!}
\frac{\rho''(z)}{R^2} + \frac{1}{3!}\frac{\rho'''(z)}{R^3}\,\,,\label{eq10}\\
\rho_w(R,\mu) \equiv \rho_w(R) = \rho_w(\infty) + \frac{\rho'(0)}{R} + 
\frac{1}{2!}\frac{\rho''(0)}{R^2} + \frac{1}{3!}\frac{\rho'''(0)}{R^3}\,\,,
\label{eq11}
\end{eqnarray}
where a prime denotes a partial derivative with respect to curvature
($R^{-1}$) and all the coefficients belong to the planar limit (at a given
chemical potential). When (\ref{eq10}) is substituted into the definition of
the adsorption $\Gamma$, in Eq.~(\ref{eq8}), we see that $\Gamma(R)$ will then
also possess an analytic curvature expansion provided the integrals over the
expansion coefficients exist (are non-divergent). In particular, the only
obvious requirement is for the existence of the planar limit integral 
\begin{displaymath}
\int_{0}^{\infty}dzz^2\left[\rho(z;\infty)-\rho_b(\mu)\right]\,\,.
\end{displaymath}
Interestingly, in the presence of an $r^{-6}$, but not $r^{-7}$, dispersion
contribution to the intermolecular potential of the solvent, this integral is
ill-defined because when an $r^{-6}$ energy is integrated over the
semi-infinite volume occupied by solvent it generates a $z^{-3}$ asymptotic
decay in the planar density profile $\rho(z;\infty)$ away from the
wall.\cite{barkerhen} Thus, one does in fact anticipate the presence of a term
or terms involving $\ln R$ in the adsorption when $r^{-6}$ dispersion
interactions are included. Indeed it has been shown explicitly that the
free-energy $\Omega_{eq}^{ex}$ then has a $\ln R$
contribution.\cite{blockbed,dietrich1} However, this non-analyticity is of
higher-order than the corresponding result for the wetting/drying case with
finite ranged potentials detailed below.\\

For a strictly finite ranged model potential one might expect the integrals
over the expansion coefficients to exist\cite{footnote} and therefore the
adsorption to possess an analytic expansion in powers of the curvature. From
the compressibility sum rule (\ref{eq8}) this property is also transferred to
the interfacial free energy (apart from unlikely terms independent of the
chemical potential). Assuming $\Gamma(R)$ is analytic in $R^{-1}$ the
right-side of the Maxwell relation (\ref{eq9}) cannot contain a term of order
$R^{-1}$, so that $\partial \rho'''(0)/\partial\mu$ in expansion (\ref{eq11})
must be zero for {\em arbitrary} values of the chemical potential
$\mu$. Stillinger and Cotter then argue that for the case of vanishing bulk
density, $\rho_b(\mu) \rightarrow 0$, $\rho'''(0)$ is zero and therefore
$\rho'''(0)=0$ for all $\mu$. If this condition is not met then one has a clear
inconsistency with (\ref{eq9}). Note that since there is no reason to expect
$\rho'''(z)$ to be zero away from the wall, a failure to appreciate the
special significance of the Maxwell relation (\ref{eq9}) would lead one to
expect a $\ln R$ term in the interfacial free
energy.\cite{jrhrevtolman,jrhjsr} In fact, the enforcement of this unexpected
condition on the limiting value of the density profile at a hard wall cavity
has recently been demonstrated numerically within density functional theory
for a hard sphere solvent.\cite{bryk} The corollary to the argument of
Stillinger and Cotter is that non-analytic terms in the curvature expansion of
the interfacial free energy demand analogous non-analytic terms to be present
in the interfacial order parameters $\rho_w$ and $\Gamma$. Although Stillinger
and Cotter did not contemplate drying at a hard cavity it is clear that
wetting/drying non-analyticity must behave consistently with their sum rule
arguments. In the following sections we provide an explicit analysis, based on
Eq.~(\ref{eq4}) and the sum rules listed above, which ascertains leading order
non-analytic behavior. 

\section{SURFACE TENSION ROUTE}
Substituting Eq.~(\ref{eq3}) into Eq.~(\ref{eq4}) we can write the interfacial
free energy as
\begin{eqnarray}
\Omega^{ex}_{eq}(R,\mu) = 4\pi R^2\left\{\gamma_{wg}(R,\mu) \right.& + 
& \gamma_{gl}(R)\left(1+ \frac{2\xi}{R}\right) + \xi\Delta\rho\delta\mu 
\nonumber \\
& + & \xi\left[\Delta\rho\delta\mu +\frac{2\gamma_{gl}^{\infty}}{R}\right]
\ln\left\{\frac{aR}{\Delta\rho\delta\mu R + 2\gamma_{gl}^{\infty}}\right\}
\left. \right\}\,\,.
\label{eq12}
\end{eqnarray}
The behavior of this quantity depends on whether $R>R_c$ or $R<R_c$, where
$R_c$ is defined by Eq.~(\ref{eq5}). There are two qualitatively different
regimes. Let us consider these two regimes separately, in each case discarding
terms beyond the leading-order curvature dependence. \\

For $R \ll R_c$, for which $\delta\mu$ must be kept sufficiently small, we
have to our desired order in the curvature expansion
\begin{equation} 
\Omega^{ex}_{eq}(R,\mu) = 4\pi R^2\left\{\gamma_{wg}(R,\mu) + 
\gamma_{gl}^{\infty} + \frac{2\xi\gamma_{gl}^{\infty}}{R}\ln\left\{
\frac{aR}{2\gamma_{gl}^{\infty}}\right\}\right\}\,\,.
\label{eq13}
\end{equation}
If we now invoke sum rule (\ref{eq7}) it follows that in this regime 
\begin{equation} 
k_BT\left[\rho_w(R,\mu)-\rho_w(\infty,\mu)\right] = \frac{2}{R}\left[
\gamma_{wg}(R,\mu) + \gamma_{gl}^{\infty}\right] + 
\frac{2\xi\gamma_{gl}^{\infty}}{R^2} \ln\left\{\frac{aR}{
2\gamma_{gl}^{\infty}}\right\}\,\,,
\label{eq14}
\end{equation}
where we remind the reader that terms of higher-order than those specified are
omitted. This is the result obtained by Evans et al. in
Ref.~\onlinecite{evans1}, based on setting $\delta\mu=0$ at the
outset.\cite{footnote2} Here, we note that for any non-zero $\delta\mu$ the
planar limit must lie outside this regime; i.e. at $R=R_c$ there must be a
cross-over to analytic behavior.\\ 

For $R \gg R_c$, an approach to the planar limit at non-zero $\delta\mu$, the
free-energy curvature expansion (\ref{eq12}) changes character to yield an
analytic form whose leading term is 
\begin{equation}
\Omega^{ex}_{eq}(R,\mu) = 4\pi R^2\left\{\gamma_{wg}(\infty,\mu) + 
\gamma_{gl}^{\infty} + \xi\Delta\rho\delta\mu\left[\ln\left\{
\frac{a}{\Delta\rho\delta\mu}\right\}+1\right]\right\}\,\,.
\label{eq15}
\end{equation}
When this result is inserted into sum rule (\ref{eq7}) one obtains 
\begin{equation}
k_BT\left[\rho_w(R,\mu)-\rho_w(\infty,\mu)\right] = \frac{2}{R}\left[
\gamma_{wg}(\infty,\mu) + \gamma_{gl}^{\infty} + \xi\Delta\rho\delta\mu\left[ 
\ln\left\{\frac{a}{\Delta\rho\delta\mu}\right\}+1\right]\right]\,\,.
\label{eq16}
\end{equation}
The next term is $O(R^{-2})$; one has an expansion analytic in
$R^{-1}$. Eq.~(\ref{eq16}) can be re-expressed as 
\begin{equation}
k_BT\left[\rho_w(R,\mu)-\rho_w(\infty,\mu)\right] = \frac{2}{R}\gamma_{wl}(\infty,\mu) 
+ O(R^{-2})\,\,,
\label{eq16b}
\end{equation}
where $\gamma_{wl}(\infty,\mu)$ is the surface tension of the planar
wall-liquid interface at chemical potential $\mu$; this quantity contains the
$\delta\mu\ln\delta\mu$ term associated with complete wetting/drying for
short-ranged forces. Thus, provided $R \gg R_c$, one can read off the planar
surface tension from the leading term in the curvature expansion of the
contact density (which is Stillinger and Cotter's route to their exact formula
for the planar surface tension discussed in Sec.~VII below) even when drying
is present.

\section{SURFACE ADSORPTION ROUTE}
When Eq.~(\ref{eq12}) is inserted into the compressibility sum rule
(\ref{eq8}) we obtain 
\begin{equation}
\Gamma(R,\mu) - \Gamma_{wg}(R,\mu)= -4\pi R^2\xi\Delta\rho
\ln\left\{\frac{aR}{\Delta\rho\delta\mu R + 2\gamma_{gl}^{\infty}}\right\}\,\,,
\label{eq17}
\end{equation}
which can be re-written as
\begin{equation}
\frac{\Gamma(R,\mu) - \Gamma_{wg}(R,\mu)}{4\pi R^2} = -\xi\Delta\rho\left[
\ln\left\{\left(\frac{1}{R} + \frac{1}{R_c}\right)^{-1}\right\}+b(T)\right]\,\,,
\label{eq17b}
\end{equation}
where $b(T)=\ln(a/2 \gamma_{gl}^\infty)$. One can expand the logarithmic term
in Eq.~(\ref{eq17b}) about $\ln R_c$ and note that the radius of convergence
(and hence the extent of the analytic regime) is $R_c/R < 1$. One can also
read off the limiting forms of the two regimes. For the non-analytic case at
$R \ll R_c$, examined by Evans et al. in Ref.~\onlinecite{evans1}, 
\begin{equation}
\Gamma(R,\mu)- \Gamma_{wg}(R,\mu) \rightarrow -4\pi R^2\xi\Delta\rho
\ln\left\{\frac{aR}{2\gamma_{gl}^{\infty}}\right\}\,\,,
\label{eq18}
\end{equation}
and for the non-zero $\delta\mu$ analytic case at $R \gg R_c$
\begin{equation}
\Gamma(R,\mu)- \Gamma_{wg}(R,\mu) \rightarrow -4\pi R^2\xi\Delta\rho
\ln\left\{\frac{a}{\Delta\rho\delta\mu}\right\}\,\,.
\label{eq19}
\end{equation}

\section{SURFACE MAXWELL RELATION ROUTE}
In this section we will check directly the consistency of our curvature
expansions for the case of drying with the surface Maxwell relation
(\ref{eq9}). For the analytic regime $R \gg R_c$ one can demonstrate
consistency by simply inserting (\ref{eq16}) into the right side of
(\ref{eq9}) and comparing with the derivative of Eq.~(\ref{eq19}). From both
routes one obtains 
\begin{equation}
\frac{\partial(\Gamma-\Gamma_{wg})}{\partial R} 
\rightarrow -8\pi R\xi\Delta\rho \ln\left\{\frac{a}{\Delta\rho\delta\mu}
\right\} \,\,. \label{eq20}
\end{equation}
The non-analytic regime $R \ll R_c$ is not so straightforward because one must 
first retain the leading order term varying with chemical potential (missing
from the right-side of Eq.~(\ref{eq14})). This requires us to keep the
non-dominant term in the prefactor of the logarithmic contribution to the
interfacial free-energy (\ref{eq12}); it is also helpful to expand the
logarithm and note that at next to leading-order it cancels the sub-dominant
$\xi\Delta\rho\delta\mu$ term:
\begin{equation}
\Omega^{ex}_{eq}(R,\mu) \rightarrow 4\pi R^2\left\{\gamma_{wg}(R,\mu) + 
\gamma_{gl}^{\infty} + \xi\left[\Delta\rho\delta\mu +
\frac{2\gamma_{gl}^{\infty}}{R}\right]
\ln\left\{\frac{aR}{2\gamma_{gl}^{\infty}}\right\}\right\}
\,\,. \label{eq21}
\end{equation}
Thus, keeping appropriate terms, at non-zero $\delta\mu$ we have
\begin{equation} 
k_BT\left[\rho_w(R,\mu)-\rho_w(\infty,\mu)\right] = \frac{2}{R}\left[
\gamma_{wg}(R,\mu) + \gamma_{gl}^{\infty} 
+ \xi\left[\Delta\rho\delta\mu+\frac{\gamma_{gl}^{\infty}}{R}\right]
\ln\left\{\frac{aR}{2\gamma_{gl}^{\infty}}\right\}\right]\,\,,
\label{eq22}
\end{equation}
where despite appearances the prefactor of the logarithmic term is not missing
a factor of two in front of $\gamma_{gl}^{\infty}$. When Eq.(\ref{eq22}) is
inserted into the surface Maxwell equation (\ref{eq9}) one immediately
rederives the result for $\partial\Gamma /\partial R$ obtained from
(\ref{eq18}), to leading order as required.

\section{RESULTS FROM DENSITY FUNCTIONAL THEORY FOR A SQUARE-WELL FLUID}
The explicit results for surface thermodynamic functions that we have
presented in earlier sections are based upon the ansatz (\ref{eq1}) for the
excess grand potential of a fluid adsorbed on a hard spherical cavity. This
ansatz, as in all interface Hamiltonian approaches to wetting/drying, relies
upon the introduction of the length $\ell$, the thickness of the drying layer
around the sphere.  In a fully microscopic approach only the fluid-fluid pair
potential should appear, once the wall-fluid (external) potential has been
specified.  The properties of the fluid are determined by the behavior of the
average one-body density $\rho(r)$, which depends, in turn, on the nature of
the wall-fluid and fluid-fluid potentials, as well as on the thermodynamic
state point ($T, \mu$) of the fluid reservoir. In order to test the
predictions of the (coarse-grained) interface Hamiltonian approach we employ
the same microscopic density functional theory (DFT) used in
Ref.~\onlinecite{evans1}. Specifically we consider an attractive square-well
fluid adsorbed at the hard spherical cavity of radius $R$.\\

The fluid-fluid potential is given by
\begin{equation}
\phi(r) = \left\{
\begin{array}{c}
\infty \,\,,\, \,\,\,\,\,\,\,\,r<\sigma\,\,\,\,\,\, \\
-\varepsilon \,\,,\, \sigma<r<3\sigma /2
\end{array}\,\,,
\right.
\label{eq24}
\end{equation}
with $\varepsilon > 0$ and the wall-fluid potential is
\begin{equation}
V(r) = \left\{                        
\begin{array}{c}
\infty \,\,,\, r<R \\
0 \,\,,\, r>R
\end{array}\,\,.
\right.
\label{eq25}
\end{equation}
The grand potential functional is taken to be
\begin{equation}
\Omega_V[\rho] = F_{id}[\rho]+F_{ex}[\rho]+\int d{\bf r}\rho({\bf r})
(V({\bf r})-\mu)\,\,,
\label{eq26}
\end{equation}
where $F_{id}[\rho]$ is the Helmholtz free energy functional of the ideal gas
and the excess (over ideal) part of the free energy functional is approximated
by
\begin{equation}
F_{ex}[\rho] = F_{hs}[\rho] + \frac{1}{2}\int d{\bf r}\int d{\bf r}'
\rho({\bf r})\rho({\bf r}') \phi_{att}(|{\bf r}-{\bf r}'|) \,\,,
\label{eq27}
\end{equation}
with $\rho({\bf r})\equiv \rho(r)$.  Here $F_{hs}[\rho]$ is the excess
Helmholtz free energy functional of a hard-sphere fluid which we treat by
means of Rosenfeld's\cite{i} successful fundamental measures
theory. Attractive interactions are treated by means of a simple mean-field
approximation;\cite{ii} we take the attractive part of the potential to be
$\phi_{att}(r) = -\varepsilon$ for $r < 3\sigma /2$ and zero otherwise.  The
equilibrium density profile $\rho_{eq}(r)$ is obtained by solving numerically
the Euler-Lagrange equation resulting from minimizing the functional
(\ref{eq26}) at fixed $\mu$, $R$, and $T$. As usual, $\Omega_V[\rho_{eq}]$
yields the estimate of the grand potential $\Omega$ and, hence, of the surface
excess quantity $\Omega_{eq}^{ex}$ defined in Eq.~(\ref{eq1}) from which other
surface thermodynamic quantities follow.  As emphasized in
Ref.~\onlinecite{evans1}, this particular DFT approach has the advantages that
i) the coexisting densities $\rho_l$ and $\rho_g$ can be calculated precisely
from the bulk free energy density arising from (\ref{eq27}), ii) the results
from the DFT defined by (\ref{eq26}) and (\ref{eq27}) satisfy the Gibbs
adsorption theorem (\ref{eq8}) and the sum rule (\ref{eq7}) for the contact
density $\rho_w(R, \mu)$, (this was confirmed earlier in extensive numerical
calculations for hard-spheres adsorbed at hard curved cavities\cite{bryk}),
and iii) the planar surface tension $\gamma_{gl}^{\infty}$ and the bulk
correlation length $\xi$ of the wetting phase (in this case gas) can be
obtained from independent calculations. Specifically, we compute
$\gamma_{gl}^{\infty}$ from a separate DFT calculation for a planar free
interface and we calculate $\xi$ by evaluating the leading-order pole of the
fluid structure factor obtained by taking two functional derivatives of
Eq.~(\ref{eq27}).\cite{iii} Note that the quantity $\xi$ entering
Eq.~(\ref{eq1}) is also the true correlation length which determines the
exponential decay of the bulk pair correlation function $g(r)$ for
$r\rightarrow\infty$.\cite{iv} The functional (\ref{eq27}) is, of course,
mean-field in character in that it omits some of the effects of capillary wave
fluctuations that occur in a wetting/drying film.\cite{ii} We shall return to
this issue later.\\

In our numerical work we chose to focus on the surface adsorption route to
surface thermodynamic functions, i.e. we chose to test the validity of
Eq.~(\ref{eq17b}) within the context of the microscopic DFT approach.  It
should be clear from the previous discussion that if the form of (\ref{eq17b})
is verified then all the relevant predictions from the coarse-grained approach
must also hold within the microscopic treatment.  Note that in
Ref.~\onlinecite{evans1} we had already confirmed, via DFT, the validity of
the predictions of the coarse-grained theory for $\delta\mu = 0$, i.e. for
$R_c = \infty$.  Here we are considering finite but large $R_c/\sigma$.  We
must also focus on the situation where $R/\sigma \gg 1$ since the mesoscopic
arguments based on (\ref{eq1}) are reliable only when the equilibrium layer
thickness $\ell_{eq}/\sigma$ is large, i.e. when $\delta\mu$ is small and the
cavity radius $R$ is large.\\ 

In Fig.~\ref{fig:gammaRc5000} we plot the difference in adsorption
$\Gamma(R,\mu)-\Gamma_{wg}(R,\mu)$, divided by the surface area $4\pi R^2$,
obtained from the DFT calculations, with $\Gamma$ defined by the integral in
Eq.~(\ref{eq8}), versus the dimensionless quantity $x$ where
\begin{equation}
x \equiv -\xi \sigma^2 \Delta\rho\ln\left\{
\frac{k_B T \sigma^{-3}}{2\gamma_{gl}^{\infty}(R^{-1} + R_c^{-1})}\right\}
\label{eq28}
\end{equation}
for fixed $R_c = 5000 \sigma$ and $k_BT/\varepsilon = 1$. The predictions of
Eq.~(\ref{eq17b}), namely that this plot should be a straight line with
positive gradient unity, are satisfied accurately both for $R \ll R_c$ and $R
\gg R_c$. We have confirmed that for $R/\sigma \geq 250$ Eq.~(\ref{eq17b})
remains valid for a wide selection of values of $R_c/\sigma$. At smaller
values of $R$ deviations of order $\sigma/R$ become apparent to the eye. We
conclude from the DFT results that provided $R/\sigma$ is sufficiently large
(and $\delta\mu$ is sufficiently small) the adsorption is given accurately by
Eq.~(\ref{eq17b}) and thus the excess grand potential is well-accounted for by
the starting equation (\ref{eq4}).\\

The results in Fig.~\ref{fig:gammaRc5000}  are appropriate to a typical
colloidal particle in `water' at 1 atmosphere. As mentioned in the
Introduction, we take $R_c = 1.4\mu$m for water at room temperature and
pressure. In Fig.~\ref{fig:gammaRc250} we chose to fix $R_c$ to be $250
\sigma$, corresponding to a thermodynamic state much further from bulk
coexistence, and varied the colloid radius $R$. The difference in adsorption
varies linearly with $x$ for large radii $R$ but deviations from a straight
line of slope unity can be discerned for smaller values of $R$.\\ 

In Fig.~\ref{fig:rhowRc250} we plot the difference between the contact density
at the spherical cavity and that a planar wall,
i.e. $\left[\rho_w(R,\mu)-\rho_w(\infty,\mu)\right]$,  versus $1/R$ for a
fixed value of $\delta\mu$ corresponding to $R_c = 250\sigma$. For $R\geq R_c$
the data lie on a straight line whose slope is given by
$2\gamma_{wl}(\infty,\mu) \sigma^2/k_B T$. Thus, the DFT results also confirm
the validity of Eq.~(\ref{eq16b}): the planar wall-liquid surface tension
$\gamma_{wl}(\infty,\mu)$ at a non-zero value of $\delta \mu$ can be obtained
from plots of the contact density versus $1/R$ provided one has sufficient
data in the range $R>R_c$. 

\section{DISCUSSION}
In this paper we have used an interface Hamiltonian appropriate to
(mean-field) wetting/drying to explore the curvature dependence of the
interfacial free energy $\Omega^{ex}_{eq}(R,\mu)$. We have concentrated on the
solvation of a hard spherical cavity of radius $R$, a problem tackled
originally by scaled particle theory, since for this system the exact sum
rules (\ref{eq7}, \ref{eq8}, \ref{eq9}) provide direct insight. In particular,
the surface Maxwell relation (\ref{eq9}) implies that terms non-analytic in
$1/R$ cannot be present in the curvature expansion of the interfacial free
energy unless there exist corresponding non-analytic terms in the Gibbs
adsorption $\Gamma(R,\mu)$ and in the contact density $\rho_w(R,\mu)$. Our
main conclusion is that in the neighborhood of complete drying one must
distinguish two regimes of behavior:  i) $R \ll R_c$, where non-analytic terms
involving $\ln R$ (see Eqs.~(\ref{eq13}, \ref{eq14}, \ref{eq18})) are present
and ii) $R \gg R_c$, where the curvature expansions involve only powers of
$1/R$. The length scale $R_c$ is given by Eq.~(\ref{eq5}); the closer the bulk
liquid is to coexistence, the smaller is the chemical potential deviation
$\delta\mu$ and the larger is $R_c$. In Ref.~\onlinecite{evans1} we considered
only the case $R_c = \infty$ ($\delta\mu =0$), identifying the leading-order,
non-analytic contributions and confirming their existence by means of DFT
calculations.   Here we elucidate the crossover between the two regimes and
verify the predictions from the interface Hamiltonian by performing DFT
calculations of the Gibbs adsorption for a wide range of $R$ and $R_c$ - see
Sec.~VI.  As the DFT yields density profiles and surface thermodynamic
functions consistent with the sum rules (\ref{eq7}-\ref{eq9}), it follows that
the leading-order curvature contributions predicted by the interface
Hamiltonian analysis are all consistent with our DFT results.\\

Of course, both theoretical approaches omit effects of capillary-wave
fluctuations. In three-dimensional systems with short-ranged wall-fluid and
fluid-fluid forces the mean-field theory of complete wetting/drying is
expected to be affected marginally by fluctuations; the upper critical
dimension is $d_c = 3$.  The linear renormalization group is sufficient to
handle these effects and amounts to a Gaussian smearing of the interfacial
binding potential.\cite{fishhuse} Since, to leading-order in $l/R$,
incorporation of curvature merely replaces $\Delta\rho\delta\mu$ for a planar
wall by $\left(\Delta\rho\delta\mu + 2\gamma_{gl}^{\infty}/R\right)$ we
conjecture,\cite{evans1} by analogy with results for the planar problem, that
our present mean-field results for the leading non-analytic contribution to
$\Omega^{ex}_{eq}(R,\mu)$, the third term on the right of Eq.~(\ref{eq13}),
should be unaltered when fluctuations are included, apart from replacing the
bulk correlation length $\xi$ by $(1 + \omega/2)\xi$, for $\omega <
2$.\cite{fishhuse} Here $\omega = k_BT/(4\pi\gamma_{gl}^{\infty}\xi^2)$ is the
usual parameter measuring the strength of capillary-wave fluctuations in
$d=3$.  Note that in the case of power-law potentials, arising from dispersion
forces, the upper critical dimension $d_c < 3$ and fluctuation effects are
not expected to alter the results of the corresponding mean-field treatment.\\

We turn now to the physical relevance of our results.  As mentioned in the
Introduction, there are potential implications for understanding aspects of
hydrophobicity at large length scales, specifically for big solvophobic solute
particles and for planar substrates.\cite{hydro} In order to appreciate some
of these we recall from the definition in Eq.~(\ref{eq1}) that the excess
chemical potential for inserting a single hard cavity into the solvent at
fixed ($\mu, T$), equivalent to the work required to create an empty cavity of
radius $R$, is given by
\begin{equation}
\mu_{hs}^{ex}(R,\mu) = p\frac{4}{3}\pi R^3 + \Omega^{ex}_{eq}(R,\mu)\,\,,
\label{eq29}
\end{equation}
where $p$ is the pressure of the reservoir. Thus, a theory for the interfacial
free energy constitutes a theory for the excess chemical potential associated
with the insertion of a hard-sphere into the solvent, precisely the quantity
that SPT attempts to calculate.  The present analysis shows that striking
logarithmic contributions can occur in $\mu_{hs}^{ex}$ for large $R$ and
sufficiently small $\delta\mu$.  It is clear that no simple extension of SPT
can hope to accommodate such subtle contributions. Indeed drying per se is not
incorporated into standard SPT treatments.\cite{jrhSPT} Moreover, it is
evident that theoretical approaches to solvation which start from an
approximate description of the Helmholtz free energy of the {\it homogeneous
  mixture} and obtain $\mu_{hs}^{ex}$ by taking the derivative with respect to
the solute density in the limit of vanishing solute will not normally be able
to capture the subtle physics associated with drying films.\\

It is likely that the physical situation for which our theory is most directly
relevant is not drying but one in which a big colloidal particle is wet by a
liquid film in the approach to liquid-gas ($\mu \rightarrow \mu_{co}^-(T)$) or
liquid-liquid coexistence.\cite{dietrich1} The arguments that lead to
Eq.~(\ref{eq1}) are valid for complete wetting by either fluid phase, provided
the wall-fluid potential has finite range or decays on a length scale shorter
than the bulk correlation length of the wetting phase. In these circumstances
the interfacial free energy per unit area will acquire a non-analytic
$R^{-1}\ln R$ contribution provided $R \ll R_c \equiv
2\gamma_{gl}^{\infty}/(\Delta\rho|\delta\mu|)$. Of course the sum rule
(\ref{eq7}) is modified when the wall-fluid potential $V_{wf}(r)$ is not purely
hard\cite{jrhrevtolman,henapp,henhen}
\begin{equation}
\frac{\partial \Omega_{eq}^{ex}(R,\mu)}{\partial R} =
-4 \pi \int_0^\infty dr r^2 \rho(r) \frac{d V_{wf}(r)}{d r} - 4 \pi R^2 p
\label{generalsumrule}
\end{equation}
However, we expect the density profile in the neighborhood of the wall to
acquire contributions equivalent to those appearing on the r.h.s. of
Eq.~(\ref{eq14}). More specifically, for a complete wetting situation, with $R
\ll R_c$, the first term on the right in Eq.~(\ref{generalsumrule}) should
acquire both a contribution $8\pi R\gamma_{gl}^{\infty}$ and a non-analytic
$8\pi\xi\gamma_{gl}^{\infty}\ln R$ contribution from the fluid-fluid interface
located near $R + \ell_{eq}$. In real fluids dispersion forces are ubiquitous
and these give rise to power-law divergences of the wetting film thickness at
a planar wall.  Analysis of the appropriate interface Hamiltonians and
extensive DFT calculations for a Lennard-Jones liquid adsorbed at a hard
spherical cavity show that wetting/drying leads to non-analytic terms in the
curvature expansion of the interfacial free energy and of the surface order
parameters that are power laws in $R^{-1}$ (not logarithms), provided $R \ll
R_c$. For example, for $R_c = \infty$, the $r^{-6}$ decay of the fluid-fluid
pair potential gives rise to the adsorption increasing as $R^{7/3}$, a term
proportional to $R^{-2/3}$ in the interfacial free energy per unit area, and a
term $R^{-5/3}$ in the contact density.\cite{evans2}\\

Throughout our discussion we have referred, somewhat loosely, to the presence
of non-analytic contributions to surface thermodynamic functions.  It is
important to understand precisely what physical repercussions such
contributions, which involve $\ln R$, might have. We choose to fix $\delta\mu$
to be small, so that drying films develop, but non-zero so that $R_c$ is very
large but finite, and vary the curvature $R^{-1}$.  For large curvature, $R
\ll R_c$, the interfacial free energy per unit area has a term in $R^{-1}\ln
R$, (see Eq.~(\ref{eq13})).  On reducing the curvature crossover occurs and
for $R \gg R_c$ the same quantity has an expansion in powers of $R^{-1}$, i.e.
\begin{equation}
\frac{\Omega^{ex}_{eq}(R,\mu)}{4\pi R^2} = \gamma_{wl}(\infty,\mu) + 
O(R^{-1})\,\,.
\label{eq30}
\end{equation}
It follows that the adsorption and the contact density differ from their
planar limiting values by terms $O(R^{-1})$ and higher powers of the
curvature; no singularity develops in derivatives as $R^{-1} \rightarrow 0^+$.
This means that one can also consider the situation of {\it negative}
curvature, $R^{-1} < 0$, corresponding to adsorption from the liquid confined
{\it inside} the hard cavity.\cite{v} (One can imagine solvent particles
`ghosted' into the cavity from a reservoir at the same fixed ($\mu, T$) - a
situation also considered in Sec.~VI of Ref.~\onlinecite{stillcott}.) By
continuity one expects thick films of gas to develop on the hard wall. Their
equilibrium thickness is still given by Eq.~(\ref{eq3}), and for the same
$\delta\mu$, films are thicker than at the planar wall since $R^{-1} < 0$.
Thus one expects a continuous (analytic) dependence of surface thermodynamic
functions on curvature in the neighborhood of $R^{-1} = 0$.  As the curvature
becomes more negative (the radius of the spherical cavity becomes smaller),
the film thickens further and eventually the denominator of the logarithm in
Eq.~(\ref{eq3}) vanishes when the curvature approaches $-R_c^{-1}$, signaling
the breakdown of the theory. However, in reality this scenario will be
prevented by the prior occurrence, at some larger radius, of capillary
evaporation, i.e. the cavity will empty leaving only the `gas phase'
\cite{evanscc}.\\

If we make $\delta\mu$ smaller the regime $R \gg R_c$, where the surface
thermodynamic functions exhibit power-law dependence on the curvature,
shrinks.  Nevertheless, provided $R_c$ remains finite there should be a
(narrow) regime where Eq.~(\ref{eq30}) and the corresponding result
(\ref{eq16b}) for the contact density remain valid. (The latter is especially
important as it forms the starting point for deriving the planar wall-liquid
tension in the analysis of Stillinger and Cotter\cite{stillcott} see
Eq.~(\ref{eq11}).)  Of course, one is free to work at bulk coexistence, $R_c =
\infty$, and then the regime where a power series expansion in the curvature
exists is of vanishing extent and $\Gamma \sim -R^2\ln R$; this is the
situation considered earlier in Ref.~\onlinecite{evans1}. In this case
capillary evaporation occurs for an infinitesimal negative curvature; the
phase transition intervenes immediately to prevent our taking the logarithm of
a negative number! Another situation one might contemplate is that in which
the radius $R$ of the hard cavity is fixed at some (large) value and the
chemical potential $\mu$ is varied.  For sufficiently large $\delta\mu$, $R
\gg R_c$ and one expects power-law curvature contributions, whereas on
approaching coexistence $\delta\mu \rightarrow 0$ and there will necessarily
be crossover to the regime $R \ll R_c$ where logarithmic contributions will
arise.\\

There are practical reasons for investigating the realm of validity of
power-series expansions in the curvature.  Helfrich's seminal analysis of
fluctuating membranes is based on the assumption that the free energy can be
written as such an expansion which includes three terms in addition to the
bulk contribution.\cite{helf,jsr} Very recently K\"{o}nig et al. \cite{vi}
have argued that the interfacial free energy per unit area,
$\Omega^{ex}_{eq}(R,\mu)/(4\pi R^2)$, and other surface thermodynamic
quantities can be written as a constant plus only two contributions, one
linear in the mean curvature and the other linear in the Gaussian curvature of
the convex surface bounding the fluid.  They support their conjecture with
numerical DFT results for a hard-sphere fluid bounded by a curved wall.  It is
clear from our analysis that the occurrence of wetting/drying negates the
possibility that power-series expansions, with a finite or an infinite number
of terms, could provide a valid description of the interfacial free energy for
every state point of the fluid.  For fluids with short-ranged wall-fluid and
fluid-fluid forces, exhibiting gas-liquid coexistence, the existence of a
large interfacial length scale $\ell_{eq} \gg \xi$ precludes such a simple
description. In the case of power-law (dispersion) forces the situation is
worse. As mentioned in Sec.~II, for $r^{-6}$ interactions $\Omega^{ex}_{eq}$ 
has a $\ln R$ contribution for a non-wet spherical
cavity.\cite{blockbed,dietrich1}\\ 

We return now to discussion of the Stillinger-Cotter paper\cite{stillcott} and
enquire whether their conclusions require revision in the light of our present
analysis.  Recall that these authors postulate the existence of a power-series
expansion (\ref{eq10}) for the density profile and on this basis they conclude
that the work of cavity formation, and therefore the interfacial free-energy,
is ``free of contributions varying logarithmically with the radius, in the
large size limit.''  As we argued in Sec.~II, their argument cannot be valid
for power-law forces.  In the case of short-ranged forces there were sound
reasons for accepting initially the validity of expansions (\ref{eq10}) and
(\ref{eq11}) but we have ascertained that the occurrence of drying may lead to
logarithmic contributions to the contact density: compare Eqs.~(\ref{eq11})
and (\ref{eq14}). Consistency with the sum rules then requires the presence of
logarithmic contributions in the interfacial free energy, in contradiction to
Stillinger and Cotter's conclusion. However, if we restrict consideration to
thermodynamic states and radii for which $R \gg R_c$ then the power-series
expansions are appropriate and Stillinger and Cotter's conclusion could be
deemed valid.  We re-iterate that these authors did not contemplate drying so
the length scale $R_c$ does not enter their analysis; effectively they set
$R_c = \sigma$ or $\xi$.\\

Another important result of Stillinger and Cotter is their formula for
$\gamma_{wl}(\infty,\mu)$, the surface tension of the {\it planar} hard-wall
liquid interface - see Eq.~(3.22) of Ref.~\onlinecite{stillcott}.  The
formula, which involves integrals over the pair correlation function of the
inhomogeneous fluid, can with hindsight be identified with exact
compressibility route expressions for the planar surface tension (discussed
below). Stillinger and Cotter obtain their formula by deriving a result for
the quantity $\rho'(0)$ entering Eq.~(\ref{eq11}). They then identify
$\rho'(0)$, the coefficient of the $R^{-1}$ term in the curvature expansion of
the contact density, with $2\gamma_{wl}(\infty,\mu)/k_BT$. This is equivalent
to employing Eq.~(\ref{eq16}) or (\ref{eq16b}).  The latter are valid provided
$R \gg R_c$.  However, this is not a handicap as one can always take the
planar limit by fixing $\delta\mu$ to be small but non-zero and allow $R^{-1}
\rightarrow 0^+$.\cite{vii}  As emphasized earlier, the resultant formula for
$\gamma_{wl}(\infty,\mu)$ does capture the gas-liquid tension plus the
$\delta\mu\ln\delta\mu$ contribution which characterizes complete drying in
systems with short-ranged forces.\\

As with much of their paper, Stillinger and Cotter's derivation of the planar 
hard wall-liquid surface tension was ahead of its time and appears to have
been missed or ignored by later workers. The subsequent history of their
result is interesting since the formula played a key role in unraveling some
of the more subtle aspects of wetting/drying.\cite{henhen,viii} Usually the
first published compressibility route expression for the planar surface
tension is attributed to Triezenberg and Zwanzig.\cite{tz} It turns out that
the earlier formula of Stillinger and Cotter is equivalent to the functional
inverse of the Triezenberg-Zwanzig (pair direct correlation function) formula,
with the latter applied to the hard wall interface rather than the free
liquid-gas interface; see for example Appendix A of
Ref.~\onlinecite{henapp}. It was not until some years later that
Schofield\cite{schof} proved the equivalence of the two compressibility route
expressions with the much earlier virial expression of Kirkwood and Buff.
Subsequently, without being aware of the work of Stillinger and Cotter, one of
us rederived their approach for the surface tension by taking curvature
derivatives of the Grand potential in order to isolate surface terms in the
curvature expansion.\cite{jrhrevtolman,henapp} A succinct general derivation,
valid for arbitrary dimensionality and wall-fluid potentials, is given in
Sec.~IIIC of Ref.~\onlinecite{henhen}.  Specializing to three-dimensions and
hard spherical cavities, the general formula reduces to
\begin{eqnarray}
\frac{1}{R}\left(\frac{\partial}{\partial R}-\frac{R}{2}
\frac{\partial^2}{\partial R^2}\right)_{\mu} \left(R^2\gamma(R,\mu)\right) 
&=& -\frac{1}{2}k_BTR^2\left(\frac{\partial\rho_w}{\partial R} 
\right)_{\mu}\nonumber \\
&=& \frac{\pi}{2}k_B T\rho_w^2\int_{0}^{2R}dr_{12}r_{12}^3\left[g_w(r_{12};R)
-1\right]\,\,,
\label{eq31}
\end{eqnarray}
where, as previously, $\gamma(R,\mu)\equiv \Omega^{ex}_{eq}(R,\mu)/(4\pi R^2)$
is the surface excess grand potential per unit area and $\rho_w\equiv
\rho_w(R,\mu)$ is the contact density of the fluid. $g_w(r_{12};R)$ is the
pair distribution function for pairs of particles 1 and 2 positioned at the
surface of the cavity.  It is clear that the exact result (\ref{eq31}) allows
one, in principle, to investigate curvature corrections to the planar tension;
only the latter was extracted by Stillinger and Cotter.  In earlier papers it
was assumed that the leading curvature correction would correspond to a Tolman
term.  Now we can see that in the presence of wetting/drying in the $R \ll
R_c$ regime, curvature corrections must generate the stronger $R^{-1}\ln R$
term in $\gamma(R,\mu)$ predicted by Eq.~(\ref{eq13}).\\

The compressibility route sum rule (\ref{eq31}) implies pronounced fluctuation
effects in the presence of drying. These are well-documented for the case of
a planar wall\cite{henhen,viii} but have not been investigated in any detail
for a curved wall.  Consider Eq.~(\ref{eq31}) in the limits where $R
\rightarrow \infty$ and $\delta\mu \rightarrow 0$, where the gas-liquid
interface will be located far from the hard wall, near $R +
\ell_{eq}$. Nevertheless, $\gamma(R,\mu)$ remains dominated by contributions
from this distant interface. The inescapable conclusion is that both the
one-body contact density $\rho_w(R\rightarrow\infty,\mu)$ and the pair
distribution, $g_w(r_{12};R\rightarrow\infty)$ contain much information about
the distant gas-liquid interface; capillary-wave like fluctuations of the
latter manifest themselves at the wall. In principle, knowledge of the
capillary-wave correlation contribution to $g_w(r_{12};R)$\cite{hensch}
would enable one to derive both the gas-liquid surface tension contribution
and the leading-order, $R^{-1}\ln R$ curvature contribution to $\gamma(R,\mu)$
predicted by the interface Hamiltonian approach. This is an ambitious
program, as previous work on the planar interface will confirm\cite{viii}!\\

There are other interesting topics discussed in the Stillinger-Cotter
paper. These include the free energy for a droplet with a fixed number of
solvent particles and some speculations on the behavior of the interfacial
free energy near the bulk critical point - see Sec.~VII of
Ref.~\onlinecite{stillcott}. Both topics warrant further attention; we cannot
do justice to these in the present paper. In concluding, however, we do wish
to return to what we believe is an important and intriguing issue in the
statistical mechanics of inhomogeneous fluids. This concerns the vanishing of
the coefficient $\rho'''(0)$ of the $R^{-3}$ term in the expansion of the
contact density $\rho_w$ - see Eq.~(\ref{eq11}). As we pointed out in Sec.~II,
Stillinger and Cotter argue, on the grounds of consistency with the sum rules,
that $\rho'''(0)$ must vanish at all thermodynamic state points\cite{reiss} -
with the caveat (see Sec.~VII of their paper) that their curvature expansions
(\ref{eq10}, \ref{eq11}) will themselves become inapplicable at or near the
gas-liquid critical point where the bulk correlation length is diverging.
Stillinger and Cotter do not write down an explicit expression for
$\rho'''(0)$ but they do write down the corresponding expression (their
Eq.~(5.13)) for $\rho_{2d}''(0)$, the coefficient of $R^{-2}$ in the expansion
of the density profile of a fluid adsorbed at a hard {\it disk} in two
dimensions.  Their sum rule arguments lead them to conclude that
$\rho_{2d}''(0)$ must be identically zero.  Commenting on this result
Stillinger and Cotter write: ``The identical vanishing of $\rho_{2d}''(0)$ in
two dimensions is hardly a transparent property of the complicated
$\rho_{2d}''(z)$ expression.''  This is certainly an understatement. If one
glances at their Eq.~(5.13) one finds integrals over combinations of one, two
and three-body distribution functions of the inhomogeneous fluid evaluated in
the planar wall limit and it is not at all obvious why $\rho_{2d}''(0)$ should
vanish. The corresponding expression for $\rho'''(0)$ in the three dimensional
fluid is even more complicated, involving additional four-body distributions.
It is remarkable that these complicated expressions should vanish identically
but this appears to be a necessary consequence of enforcing consistency with
the Maxwell relation (\ref{eq9}).\\

What then is the status of Stillinger and Cotter's conclusions regarding the
vanishing of $\rho'''(0)$ and $\rho_{2d}''(0)$? Their arguments require the
existence of the curvature expansion (\ref{eq10}).  This is invalid for
power-law intermolecular potentials and must fail near the critical point for
any type of interaction. We have shown that the expansion can also fail, owing
to the onset of drying, for all states sufficiently close to gas-liquid
coexistence.  Given the existence of these exceptions, one might be concerned
that the expressions derived by Stillinger and Cotter for $\rho'''(0)$ and
$\rho_{2d}''(0)$ in terms of integrals over correlation functions might not
necessarily exist for all thermodynamic state points. However, one could
restrict attention to states where the bulk correlation length $\xi \ll R$ and
where wetting/drying does not occur or to states where $R > R_c$ if
wetting/drying does occur. In fact, this latter restriction is implied in
footnote~8 of a slightly later paper by Stillinger\cite{still73} - another
paper remarkable for its allusions to future developments; in particular,
Stillinger almost arrives at a modern statement of drying (and hence wetting)
transitions (see his Fig.~3 and footnote~5). Under these restrictions the
expansions (\ref{eq10}, \ref{eq11}) should exist. Do we then accept that the
Stillinger-Cotter sum rule, their Eq.~(5.13), is ill-defined outside these
restrictions? This is feasible because the right-side of their complicated
compressibility route sum rule contains integrals over the planar distribution
functions alone, which could become very long-ranged in the presence of
capillary-wave fluctuations (or bulk critical fluctuations or power-law
interactions).\\

However, provided $\delta \mu >0$, it is difficult to see why the sum of
integrals should vanish for one state and then not exist for another close
by. The correlation length for complete wetting diverges only at $\delta \mu
=0$. What is more likely is that the sum of integrals is zero for all state
points with $\delta \mu >0$ but in the non-analytic $R<R_c$ regime, where the
power series expansion is not valid, there could be an additional,
non-vanishing contribution to the $R^{-3}$ term in (\ref{eq11}) that implies a
logarithmic term in the free energy.\cite{ix}\\

These matters are not of purely academic interest. There are many practical
applications, see e.g. Ref.~\onlinecite{vi}, where it would be beneficial to
have a theory that expresses the interfacial free energy and density profiles
as power series expansions in the curvature. As mentioned earlier, DFT studies
of hard-sphere fluids\cite{bryk} provide compelling numerical evidence for the
vanishing of $\rho'''(0)$. Those studies are based on the Rosenfeld
functional\cite{i} which has its basis in scaled particle theory, i.e. it is
constructed from a finite number of fundamental (geometric) measures. However,
it should not be the particular form of the Rosenfeld functional that dictates
$\rho'''(0)=0$. {\em Any} DFT satisfying the sum rules (\ref{eq7}-\ref{eq9})
must necessarily yield $\rho'''(0)=0$ at a hard cavity, provided the curvature
expansion exists (in particular, $R>R_c$). If follows that the DFT must yield
distribution functions that are consistent with the Stillinger-Cotter sum
rule. The behavior of $\rho'''(0)$ defined by our DFT applied to the
non-analytic regime $R<R_c$ requires further analysis. \\\\

\noindent ACKNOWLEDGMENTS:  
We have benefited from discussions with P. K\"{o}nig, K. Mecke, A. O. Parry,
J. Stecki and especially M. Thomas. RE is grateful to S. Dietrich and his
colleagues for kind hospitality and to the Humboldt Foundation for support
under GRO/1072637 during his visits to Stuttgart.

\newpage

\begin{figure}[t]
\caption{\label{fig:gammaRc5000}The difference in adsorption $\Gamma(R,\mu) -
\Gamma_{wg}(R,\mu)$, divided by $4 \pi R^2$, for a square-well fluid in the
liquid and gas phases versus $x=-\xi \sigma^2 \Delta \rho \ln \{k_B T
\sigma^{-3} /(2 \gamma_{gl}^\infty(R^{-1}+R_c^{-1})) \}$ for fixed temperature
$k_B T/\varepsilon=1$ and fixed $R_c\equiv 2 \gamma_{gl}^{\infty}/(\Delta \rho
\delta\mu)=5000 \sigma$, a value typical for water. The results of the DFT
calculation (symbols and line) are for sphere radii ranging between $R=5\cdot
10^5 \sigma$ (bottom) and $R=295 \sigma$ (top). The line has gradient unity
consistent with the prediction of Eq.~(\ref{eq17b}). For this temperature
$\Delta \rho \sigma^3 = 0.54542$, $\gamma_{gl}^\infty \sigma^2/k_B T =
0.19327$ and the correlation length of the coexisting gas phase $\xi=0.48962
\sigma$.} 
\end{figure}

\begin{figure}[t]
\caption{\label{fig:gammaRc250} As in Fig.~\ref{fig:gammaRc5000} but now for
fixed $R_c=250 \sigma$. The sphere radii range between $R=5\cdot 10^5 \sigma$
(bottom) and $R=295 \sigma$ (top). Small deviations from a straight line with
gradient unity can be ascertained at small values of $R$ (large $x$).}
\end{figure}

\begin{figure}[t]
\caption{\label{fig:rhowRc250} Difference between the contact density at the
spherical cavity and that at a planar wall,
$\rho_w(R,\mu)-\rho_w(\infty,\mu)$, versus $\sigma/R$ for a fixed $R_c=250
\sigma$ and $k_B T/\varepsilon = 1$. The solid line denotes the DFT
results. For $R\gtrsim R_c$ these lie on the straight (dotted) line whose
slope is given by the planar tension $2 \gamma_{gl}^\infty(\infty,\mu)
\sigma^2/k_B T$, as predicted by Eq.~(\ref{eq16b}).}
\end{figure}
\clearpage 

\begin{figure}
\centering\epsfig{file=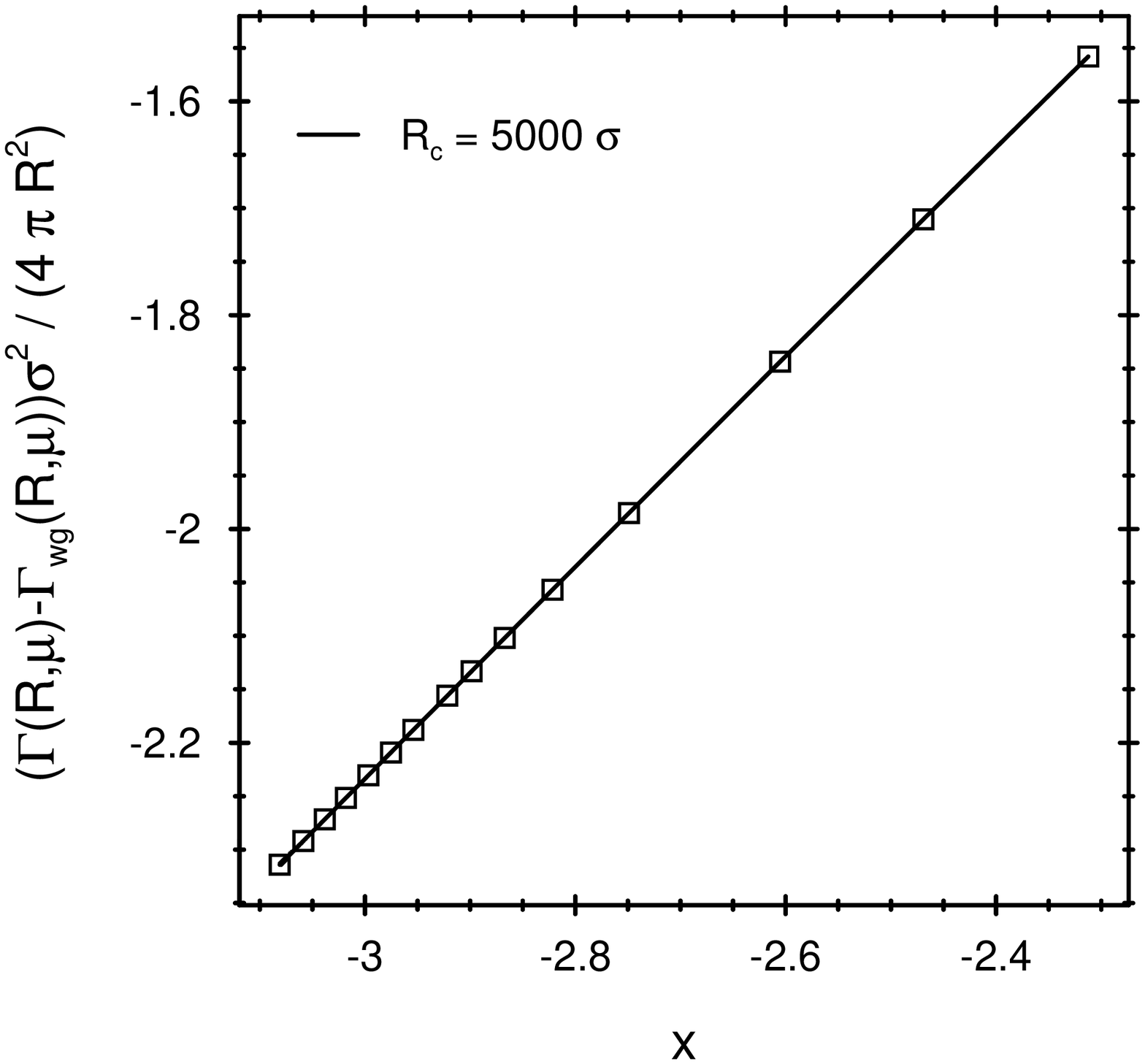,width=10cm}
\vspace{0.5cm}
\\
R. Evans, J.R. Henderson, and R. Roth, Fig.~\ref{fig:gammaRc5000}
\end{figure}
\clearpage

\begin{figure}
\centering\epsfig{file=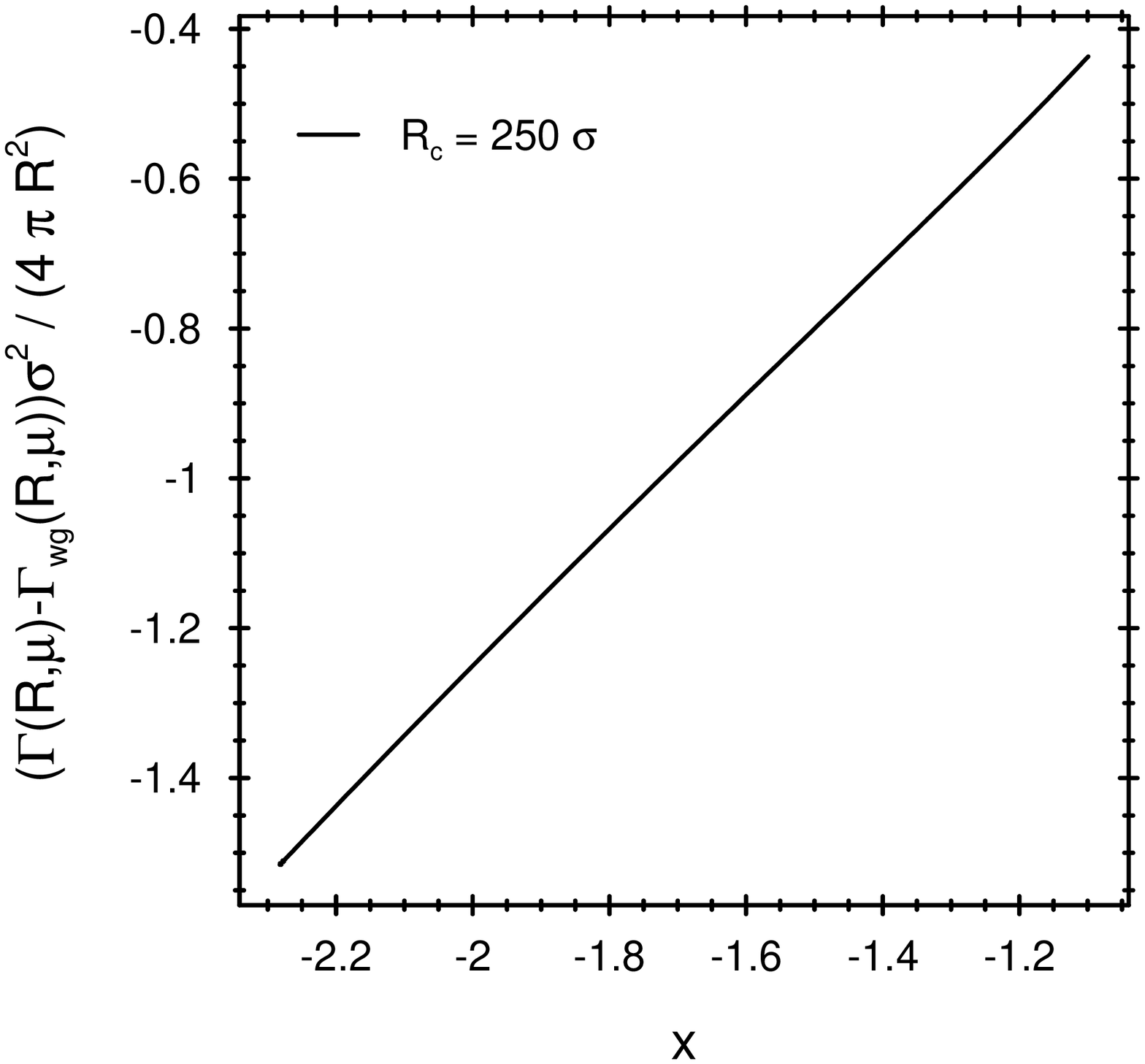,width=10cm}
\vspace{0.5cm}
\\
R. Evans, J.R. Henderson, and R. Roth, Fig.~\ref{fig:gammaRc250}
\end{figure}
\clearpage

\begin{figure}
\centering\epsfig{file=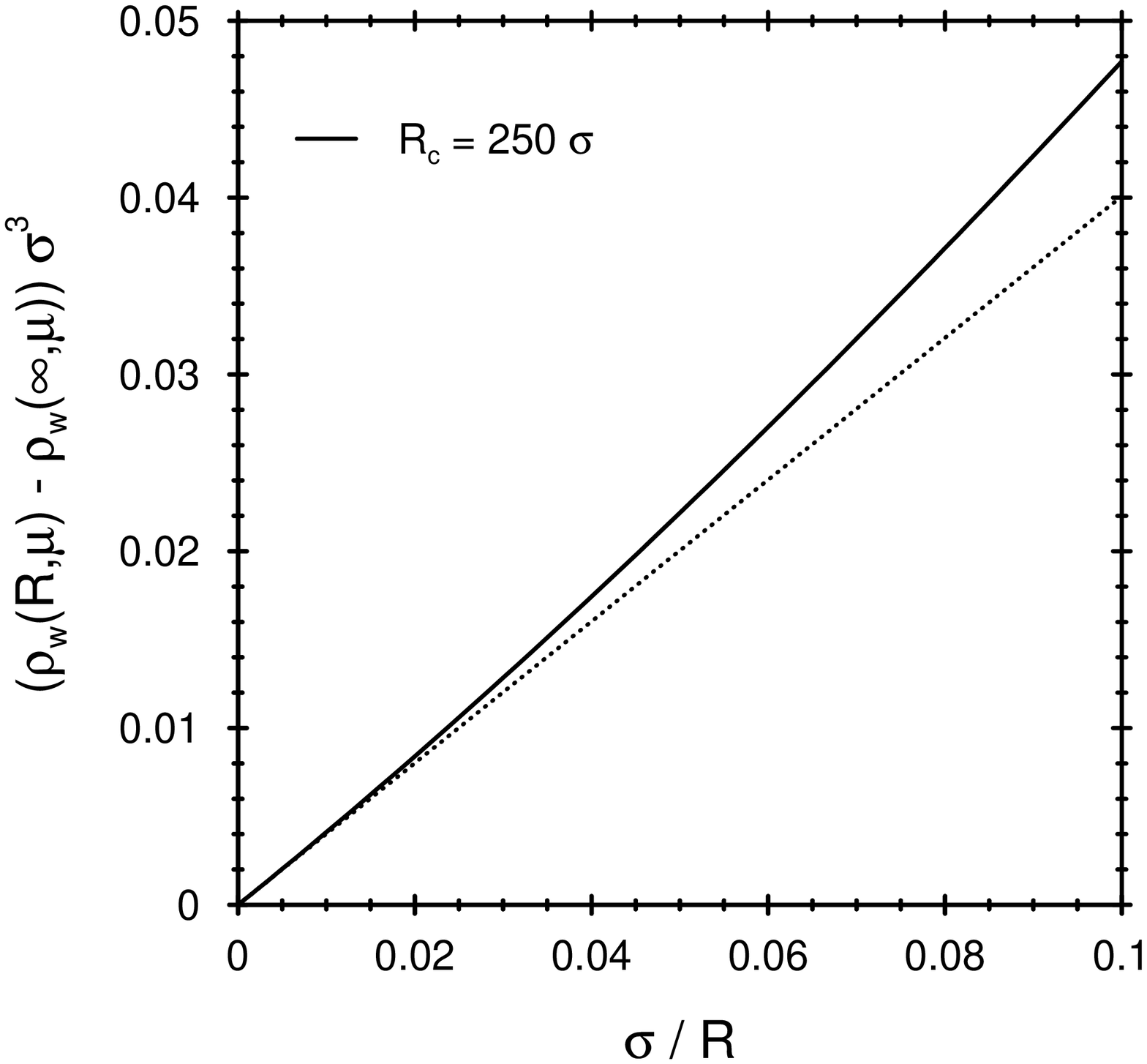,width=10cm}
\vspace{0.5cm}
\\
R. Evans, J.R. Henderson, and R. Roth, Fig.~\ref{fig:rhowRc250}
\end{figure}
\clearpage


\begin{thebibliography}{99}
\bibitem{footnote} Note that for finite ranged potentials even in a drying
situation, where a macroscopically thick film of gas forms at the planar wall,
the integral $\int_{0}^{\infty}dzz^2\left[\rho(z;\infty)-\rho_b(\mu)\right]$
is finite provided $\delta\mu > 0$.
\bibitem{footnote2} The quantity $\xi \gamma_{gl}^\infty/k_B T$, which has the
dimension of an inverse length, should determine the amplitude of logarithmic
contributions to the excess free energy for any wetting/drying situation in a
fluid with short-ranged forces. In oral presentations of this research we have
referred to $\xi \gamma_{gl}^\infty/k_B T$ as the Paulaner coefficient after
the institute `Zum Paulaner' where its significance was first recognized. Away
from the bulk critical point $\xi < \sigma$ and $\gamma_{gl}^\infty\lesssim k_B
T/\sigma^2$, where $\sigma$ is an atomic diameter, and the Paulaner coefficient
is considerably smaller than $\sigma^{-1}$. For the square-well fluid
investigated in Sec.~VI the Paulaner coefficient takes the value
$0.0946\sigma^{-1}$ at $k_B T/\varepsilon = 1$.
\bibitem{v} We should note that the sum rule (\ref{eq9}) also pertains to
fluid adsorbed inside the cavity. $\rho_w(R, \mu)$ is the contact density on
the interior of the wall and the adsorption is now defined as $\Gamma(R,\mu)
\equiv 4\pi \int_{0}^{R}drr^2\left[\rho(r)-\rho_b(\mu)\right]$. Note also that
(\ref{eq9}) is a close analog of the Maxwell relation
\begin{equation}
\frac{\partial \Gamma(L,\mu)}{\partial L} = k_BT\frac{\partial}{\partial \mu}
\left[\rho_w(L,\mu)-\rho_w(\infty,\mu)\right] = \frac{\partial f_s(L,\mu)}{\partial\mu}\,\,,
\nonumber
\end{equation}
valid for a fluid confined by two parallel, planar hard walls separated by a
distance $L$.  $\Gamma(L,\mu)$ is the Gibbs adsorption per unit area and
$f_s(L,\mu)$ is the so-called solvation force (or disjoining pressure),
i.e. the excess pressure arising from confinement.\cite{henhen}  One can
obtain the adsorption by integrating $k_BT\frac{\partial}{\partial \mu}
\left[\rho_w(L,\mu)-\rho_w(\infty,\mu)\right]$ w.r.t. $L$ just as one can
obtain $\Gamma(R,\mu)$ by integrating the r.h.s. of (\ref{eq9}) w.r.t. $R$.
\bibitem{vii} Note that caution must be exercised with the order of the
limits.  For example, using sum rule (\ref{eq9}) with Eq.~(\ref{eq16}) leads
to the prediction $\Gamma(R,\mu) \sim R^2\ln\delta\mu$, which would seem to
imply an unphysical divergence of the adsorption (at finite $R$) as $\delta\mu
\rightarrow 0$. However, Eq.~(\ref{eq16}) is valid only for $R \gg R_c$,
i.e. one requires $\delta\mu > 2\gamma_{gl}^{\infty}/(R\Delta\rho)$.
\bibitem{ix} In the case of a hard cylinder of radius $R$ the term in question
is known to be $\xi\gamma_{gl}^{\infty}/(k_BTR^2)$, for $R<R_c$, which is
certainly non-vanishing - see Eq.~(8) of Ref.~\onlinecite{evans1}. This does
not, of course, prove that the right-side of Stillinger and Cotter's sum rule
Eq.~(5.13) remains well-defined. Note that for the sphere a $R^{-3}$ term in
the expansion of $\rho_w(R)$ implies a contribution proportional to $\ln R$ in
$\Omega_{eq}^{ex}$. This would be weaker than the $R \ln R$ contribution
associated with complete drying - see present Eq.~(\ref{eq13}).
\end{thebibliography}
\end{document}